\documentclass[10pt]{article}

\usepackage{amsmath}

\usepackage{array}

\usepackage{appendix}

\usepackage{tocloft}                   

\usepackage{graphicx}

\usepackage{amsfonts}

\usepackage{amssymb}

\usepackage{mathrsfs}

\usepackage{yfonts}

\usepackage{euscript}

\usepackage{centernot}                 

\usepackage{ifsym}                     

\usepackage{lmodern}                   

\usepackage{upgreek}

\usepackage{mathtools}

\usepackage{color}

\usepackage{slantsc}
\usepackage{calligra}

\usepackage{bbold}          

\usepackage[T1]{fontenc}

\usepackage{epsf}

\usepackage{latexsym}

\usepackage{tipa}

\usepackage{makeidx}

\makeindex

\def\b{\begin{equation}}

\def\e{\begin{equation}}

\def\be{\begin{equation}}              

\def\ee{\end{equation}}

\def\beq{\begin{equation}}

\def\eeq{\end{equation}}

\def\bea{\begin{eqnarray}}

\def\eea{\end{eqnarray}}

\def\m{\mbox{ }}

\def\mma {\m , \m \m }

\def\!{\hspace{-1.6667em}}

\def\n{\noindent}

\def\u{\underline}

\def\slLambda{\mathit{\Lambda}}                   

\def\bia{\mbox{\boldmath$a$}}

\def\bix{\mbox{\boldmath$x$}}

\def\biy{\mbox{\boldmath$y$}}

\def\mF{\mbox{F}}

\def\mI{\mbox{I}}                        

\def\mJ{\mbox{J}}  

\def\mK{\mbox{K}}

\def\mL{\mbox{L}}

\def\mM{\mbox{M}}                        

\def\mN{\mbox{N}}

\def\mU{\mbox{U}}                        

\def\mh{\mbox{h}}

\def\mp{\mbox{p}}

\def\ms{\mbox{s}}

\def\bh{\u{\u{\mbox{h}}}  }            

\def\bM{\mbox{\bf M}}

\def\bN{\mbox{\bf N}}

\def\bh{\mbox{\bf h}}

\def\bp{\mbox{\bf p}}

\def\bupchi{\mbox{\boldmath$\chi$}}                     

\def\bupiota{\mbox{\boldmath$\iota$}}                   

\def\bcalC{\mbox{\boldmath ${\cal C}$}}

\def\bcalK{\mbox{\boldmath ${\cal K}$}} 

\def\bcalD{\mbox{\boldmath ${\cal D}$}}

\def\scM{\mbox{\scriptsize ${\cal M}$}}                    

\def\fz{\mbox{\sffamily z}}

\def\sa{\mbox{\scriptsize a}}

\def\se{\mbox{\scriptsize e}}

\def\sg{\mbox{\scriptsize g}}

\def\si{\mbox{\scriptsize i}}

\def\sll{\mbox{\scriptsize l}}  

\def\sm{\mbox{\scriptsize m}}

\def\sn{\mbox{\scriptsize n}}

\def\sr{\mbox{\scriptsize r}}

\def\st{\mbox{\scriptsize t}}

\def\sv{\mbox{\scriptsize v}}

\def\sx{\mbox{\scriptsize x}}

\def\sF{\mbox{\scriptsize F}}

\def\sL{\mbox{\scriptsize L}}

\def\sfz{\mbox{\sffamily{\scriptsize z}}}     

\def\sfA{\mbox{\sffamily{\scriptsize A}}}     

\def\sfZ{\mbox{\sffamily{\scriptsize Z}}}      

\def\sbN{\mbox{{\bf \scriptsize N}}}

\def\sbcS{\mbox{\boldmath \scriptsize ${\cal S}$}}

\def\bscU{\mbox{{\boldmath \scriptsize${\cal U}$}}}                               

\def\bscS{\mbox{\boldmath \scriptsize${\cal S}$}}                               

\def\bigupsigma{\mbox{\Large$\sigma$}}

\def\Thomas{\,\,\mbox{\textcircled{$\rightarrow$}}\,\,}

\def\sumi2{\sum\mbox{}_{\mbox{}_{\mbox{\scriptsize $i$=1}}}^2}

\def\sumi3{\sum\mbox{}_{\mbox{}_{\mbox{\scriptsize $i$=1}}}^3}

\def\sumABcycles3{\sum\mbox{}_{\mbox{}_{\mbox{\scriptsize cycles $A,B$=1}}}^{3}}

\def\sumCDcycles3{\sum\mbox{}_{\mbox{}_{\mbox{\scriptsize cycles $C,D$=1}}}^{3}}

\def\sumj3{\sum\mbox{}_{\mbox{}_{\mbox{\scriptsize $j$=1}}}^3}

\def\sumk3{\sum\mbox{}_{\mbox{}_{\mbox{\scriptsize $k$=1}}}^3}

\def\prodiA1{\prod\mbox{}_{\mbox{}_{\mbox{\scriptsize $i$=1}}}^{A - 1}}

\def\sumpsi{\sum\mbox{}_{\mbox{}_{\mbox{\scriptsize $\uppsi$}}}}                              

\def\d{\textrm{d}}                                                  

\def\pa{\partial}                                                   

\def\bpa{\mbox{\boldmath$\partial$}}                                                   

\def\es{\m = \m}

\def\:={\m := \m}

\def\=:{\m =: \m}

\def\lFrg{\mbox{\Large$\mathfrak{g}$}}                         
\def\nFrg{\mbox{\large$\mathfrak{g}$}}                         

\def\Hilb{\mbox{{\boldmath$\mathfrak{H}$}ilb}}                 
\def\lt{\mbox{\Large $t$}}                                 

\def\scD{\mbox{\scriptsize ${\cal D}$}}                    

\def\scH{\mbox{\scriptsize ${\cal H}$}}                    

\def\bscS{\mbox{\boldmath\scriptsize ${\cal S}$}}

\def\scS{\mbox{\scriptsize ${\cal S}$}}                    

\def\FrQ{\mbox{\Large $\mathfrak{q}$}}                               
\def\Phase{\mbox{{\boldmath$\mathfrak{P}$}hase}}                     

\def\bFrR{\mbox{\boldmath$\mathfrak{R}$}}                            
\def\Rig-Phase{\bFrR\mbox{ig-}\Phase}                                
                                                                													   
\def\bFrR{\mbox{\boldmath$\mathfrak{R}$}}                            

\def\bFrR{\mbox{\boldmath$\mathfrak{R}$}}                            

\def\1mat{\u{\u{1}}}                                                 

\def\Positive-Modespace{\mbox{{\boldmath$\mathfrak{M}$}odespace$^+$}}

\def\POSITIVE-MODESPACE{\mbox{{\boldmath$\mathfrak{M}$}ODESPACE$^+$}}
                                                                                                                             														
\def\Riem{\bFrR\mbox{iem}}                                           

\def\Kin-Hilb{\mbox{{\boldmath$\mathfrak{K}$}in-\Hilb}}                     

\def\Mid-Hilb{\mbox{{\boldmath$\mathfrak{M}$}id-\Hilb}}                     

\def\Dyn-Hilb{\mbox{{\boldmath$\mathfrak{D}$}yn-\Hilb}}                     

\def\5Star{\mbox{\Large$\star$}}              

\def\K{Kucha\v{r} }

\begin{document}

\begin{center}

\Huge{\bf A LOCAL RESOLUTION OF}

\vspace{.1in}

\normalsize

\Huge{\bf THE PROBLEM OF TIME}

\vspace{.1in}

\Large{\bf IX. Spacetime Constructability}

\vspace{.15in}

{\large \bf E. Anderson} 

\vspace{.15in}

{\large \it based on calculations done at Queen Mary, University of London;}

\vspace{.15in}

{\large \it Peterhouse, Cambridge; }

\vspace{.15in}

{\large \it Universit\'{e} Paris VII.  }

\end{center}

\begin{abstract}

Assuming Temporal and Configurational Relationalism, 
GR as Geometrodynamic's DeWitt supermetric alongside local Lorentzian Relativity with its universal finite maximum propagation speed 
arises as one of very few options from Feeding Families through a Dirac-type Algorithm for Consistency. 
This amounts to Spacetime Construction from prior assumptions about space and dynamics alone.  
The other alternatives, arising as cofactors' likewise strongly vanishing roots are, 
firstly, Galileo-Riemann Geometrostatics with Galilean Relativity's infinte propagation speed. 
Secondly, Strong Gravity with Carrollian Relativity 's zero propagation speed.  
If none of these vanish, constant mean curvature of the spatial slice is enforced, 
paralleling previous work on decouping GR's constraints and addressing its initial-value problem. 
Assuming just Temporal Relationalism, spatial 3-diffeomorphism Configurational Relationalism is enforced as an integrability as one of very few options. 
The alternatives here are, once again, Galileo-Riemann Geometrostatics and Strong Gravity, for which spatial 3-diffeomorphisms are thereby optional, 
and local volume preserving diffeomorphisms.
Options are few in each case above due to Rigidity Kicking In.  

\m 

We furthermore demonstrate that such Rigidities are more generally Lie rather than specific to Dirac-type Algorithms. 
We do this by deriving the conformal versus projective ambiguity in top-geometry by feeding the general quadratic generator into the Lie Algorithm for Consistency. 
Both Feeding Families through a Brackets Consistency Algorithm and Rigidity are thus additionally of relevance to the Foundations of Geometry.

\end{abstract}

\section{Introduction}

This is the ninth Article \cite{I, II, III, IV, V, VI, VII, VIII} on the Problem of Time 
\cite{Battelle, DeWitt67, Dirac, K81, K91, K92, I93, K99, APoT, FileR, APoT2, AObs, APoT3, ALett, ABook, A-CBI} and its underlying Background Independence. 
Herein, we combine the following structures and techniques. 

\m

\n a) Articles V and VI's on the consistent composition of thes two Relationalisms presented piecemeal in Articles I and II. 

\m 

\n b) Article VII's subsequent composition of these with Constraint Closure via the TRi Dirac-type Algorithm.  

\m 

\n c) Article III's piecemeal approach to Spacetime Construction \cite{RWR, Than, Phan, Lan2}.  

\m 

\n All in all, we are Feeding Families of Theories (Sec 2), 
from which constraints are provided by Temporal and Configurational Relationalism, 
through the TRi-Dirac-type Algorithm \cite{AM13}, which succeeds by Rigidity Kicking In. 
This suceess (Sec 3) is in the form of structural features of GR-as-Geometrodynamics -- GR's specific configuration space metric --
alongside localy-Lorentzian relativity universal finite maximum propagation speed  
being recovered as one of a very small number of alternatives realized as algebraic roots of an equation arising from this Dirac-type Algorithm.  
I.e.\ a strongly-vanishing equation with two other factors; the alternatives thus encoded are as follows (Secs 4 and 5).  

\m 

\n I) Galileo--Riemann Geometrostatics \cite{Phan, AM13} alongside locally-Galilean Relativity with its universal finite maximum propagation speed.  

\m 

\n II) Strong-Gravity-Geometrodynamics \cite{San, Phan, AM13} alongside locally-Carrollian Relativity \cite{LL65, BL68}.

\m 

\n Thereby, these three alternatives constitute the finite-infinite-zero universal fundamental propagation speed ('of light') trilemma: 
the eventual and logically-complete extension of Einstein's Lorentzian versus Galilean Relativity dilemma (Sec 6). 
This is now moreover realized as an equation that GR's constraints plus Dirac's mathematics {\sl gives}, 
so it arrived at routinely, i.e.\  without needing an Einstein to intuit it. 

\m 

\n If none of these vanish, constant mean curvature of the spatial slice is enforced (Sec 7), 
paralleling previous work on decouping GR's constraints and addressing its initial-value problem. 

\m 

\n Sec 8 covers the `Discover and Encode' approach to Physics, using Metrodynamics as an example.
Here just Temporal Relationalism is assumed, 
with spatial 3-diffeomorphism Configurational Relationalism now enforced as an integrability as one of very few options. 
The alternatives here are, once again, Galileo-Riemann Geometrostatics and Strong Gravity, for which spatial 3-diffeomorphisms are thereby optional, 
and local volume preserving diffeomorphisms.

\m 

\n We finally demonstrate in Sec 9 that such Rigidities are more generally Lie \cite{Lie80, Lie} rather than specific to Dirac-type Algorithms \cite{Dirac}. 
We attain this \cite{A-Brackets} by deriving the conformal versus projective ambiguity in top-geometry 
by feeding the general quadratic generator into the Lie Algorithm for Consistency. 
In this way, Feeding Families through a Brackets Consistency Algorithm, and Rigidity, 
are also of further relevance to the Foundations of Geometry \cite{HC32, Stillwell, PE-1, A-Brackets} itself.  
This broader `Lie magic' rather than just `Dirac magic' observation finally paves 
the way for quantum-commutator Lie brackets analogues of Spacetime Construction: a key step in the quantum-level Problem of Time.

\section{Relational first principles ansatz for geometrodynamical theories}

We begin with the usual choice of $\FrQ = \Riem(\bigupsigma)$ and $\lFrg = Diff(\bigupsigma)$. 

\m 

\n{\bf Structure 1} We however now entertain a more general ansatz for a family of candidate geometrical actions 
built from differentiable and metric level spatial objects \cite{RWR, FileR, AM13}, 
\be
{\cal S}_{w,y,a,b}  \es  {\cal S}_{w,y,a,b}  
                    \es  \iint_{\bigupsigma} \d^3x \sqrt{\overline{a \, {\cal R} + b}} \, \pa \ms_{w,y}  \m . 
\label{trial} 
\ee
The line element $\pa \ms_{w,y}$ here is built out, firstly, the usual 
\be 
\pa_{\underline{\sF}}  \:=  \pa - \pounds_{\pa\u{\sF}}   \m .
\ee
Secondly, the more general if still ultralocal -- $\bpa \bh$-independent -- supermetric $\bM_{w,y}$ kinetic metric is homogeneous quadratic in the changes, 
\be 
\bM_{w, x} \m \mbox{ with components } \m  
\mM^{abcd}_{w,y} \:=  \frac{1}{y} \{\mh^{ac} \mh^{bd} - w \, \mh^{ab} \mh^{cd}\}                      \m .
\ee 
Its densitized inverse is 
\be
\bN^{x,y}         \m \mbox{ with components } \m 
\mN_{abcd}^{x,y}               \:=                 \frac{y}{\sqrt{h}}\left\{  \mh_{ac} \mh_{bd}  -  \frac{x}{2} \,\mh_{ab} \mh_{cd} \right\}        \m , 
\ee 
for  
\be 
x \:= \frac{2 \, w}{3 \, w - 1}                                                                                                                     \m .   
\ee
The parametrization by $x$ is chosen for GR to be the 
\be 
w = x = y = 1
\ee 
case. 

\m 

\n{\bf Remark 1} To maintain $\bM$'s invertibility, we exclude the case 
\be 
w = \frac{1}{3}  \m ;
\ee 
we already noted this value's pathologicalness in Article VI.

\m 
 
\n{\bf Structure 2} The conjugate momenta are then 
\be
\bp  \es  \bM_{w,y} \cdot \frac{\pa_{\underline{\sF}} \bh}{2 \, \pa\mI}  \m ; 
\ee
these generalize the relational version (VI.64) of GR's ADM momenta \cite{ADM} (II.18).

\m 

\n{\bf Structure 3} The quadratic primary constraint encoded by this action is 
\be
\scH^{\st\sr\si\sa\sll}  \es  \scH_{x,y,a,b} 
                         \:=  {||\bp||_{\sbN^{x, y}}}^2        - \overline{a \, {\cal R} + b}   
                         \:=  \mN_{abcd}^{x,y}\mp^{ab}\mp^{cd} - \overline{a \, {\cal R} + b} 
						 \es  0                                                                                                                   \m . 
\label{H-trial}
\ee
\n{\bf Structure 4} The secondary constraint is just the usual GR momentum constraint $\u{\scM}$.

\section{Geometrodynamical consistency, local Relativity and Spacetime Construction}

\n{\bf Consistent Geometrodynamics Theorem}. 
If the geometrodynamical ansatz (\ref{trial}) is assumed, the following four outcomes are consistent.  

\m 

\n i)  Recovery of GR. 

\m 

\n ii) A 1-parameter family of geometrostatics.

\m 

\n iii) A 1-parameter family of strong gravity theories. 

\m 

\n iv) A group of formulations and theories based upon 
\be
{\cal D}_i  \left\{ \frac{\mp}{\sqrt{\mh}} \right\} = 0 \m .
\ee  
\n {\bf Consistent Relativities Theorem}.  
Upon adding minimally-coupled matter, emergent local Relativity is Lorentzian for i), Galilean for ii) and Carrollian for iii).
This is in the sense of an emergent shared propagation speed that is finite for i), infinite for ii) and zero for iii).

\m

\n {\bf Classical Spacetime Construction Theorem}.  
In case i), GR spacetime emerges by construction from assuming of just space, Temporal and Configurational Relationalism.

\m

\n Toward establishing these theorems \cite{RWR, San, ABFKO, AM13}, form the Poisson brackets of the constraints and apply the TRi Dirac-type Algorithm.  
This gives \cite{AM13} (III.53-54) with the family of constraints $\scH_{x,y,a,b}$ in place of $\scH$, alongside  
$$ 
\mbox{\bf \{} (    \scH_{x,y,a,b}    \, | \,    \pa \mJ    ) \mbox{\bf , } \, (    \scH_{x,y,a,b}    \, | \,    \pa \mK    ) \mbox{\bf \}}  = 
- 2 a \, y \, (    \u{\bcalD} \u{\u{\bp}}\mbox{}    +      \{x - 1\}     \u{\bcalD} \,  \mp \, | \, {\pa \mJ} \, \overleftrightarrow{\u{\bpa}} {\pa \mK}   )  =
$$
\be
    a \, y \, (    \u{\scM}                  + 2           \{1 - x\}     \u{\bcalD} \mp \, | \, {\pa \mJ} \, \overleftrightarrow{\u{\bpa}} {\pa \mK}   )
=   a \, y \, (    \u{\scM}      \, | \,    \pa \mJ \, \overleftrightarrow{\u{\bpa}} \pa \mK    ) 
                                             + 2 \, a \, y \{1 - x\} (   \u{\bcalD} \, \mp \, | \, {\pa \mJ} \, \overleftrightarrow{\u{\bpa}} {\pa \mK}   )  .
\label{Htrial-Htrial}
\ee
[C.f.\ the Dirac algebroid of GR \cite{Dirac51, Dirac58, Dirac} as a well-known subcase.]   

\m 

\n The first equality is a lengthy brackets evaluation. 
The second recognizes the presence of $\u{\scM}$.
The third expands out the terms, to reveal that the habitual GR piece
\be
a \, y \, (    \u{\scM}  \cdot  \u{\u{\bh}}^{-1}  \cdot    \, | \,    \pa \mJ \, \overleftrightarrow{\u{\bpa}} \pa \mK    ) 
\label{Int}
\ee
has additively picked up a {\it ne constraint, specifier or obstruction term} \cite{AM13} with four factors:  
\be
2 \times a \times y \times \{ 1 - x \} \times (    \bcalD \mp \, | \, \pa \mJ \, \overleftrightarrow{\bpa} \pa \mK    ) \m .  
\label{4-Factors}
\ee
Each factor here provides a different way in which to complete the TRi Dirac-type Algorithm. 

\m 

\n The first three are strongly vanishing options (Sec \ref{3-Strong}), 
whereas the fourth is a weakly vanishing option which includes cases in which the TRi Dirac-type Algorithm has further steps (Sec \ref{1-weak}).  
Any of these options give automatic closure, so $\scH_{x,y,a,b}$ is rendered first-class.
%
{\begin{figure}[!ht]
\centering
\includegraphics[width=1.0\textwidth]{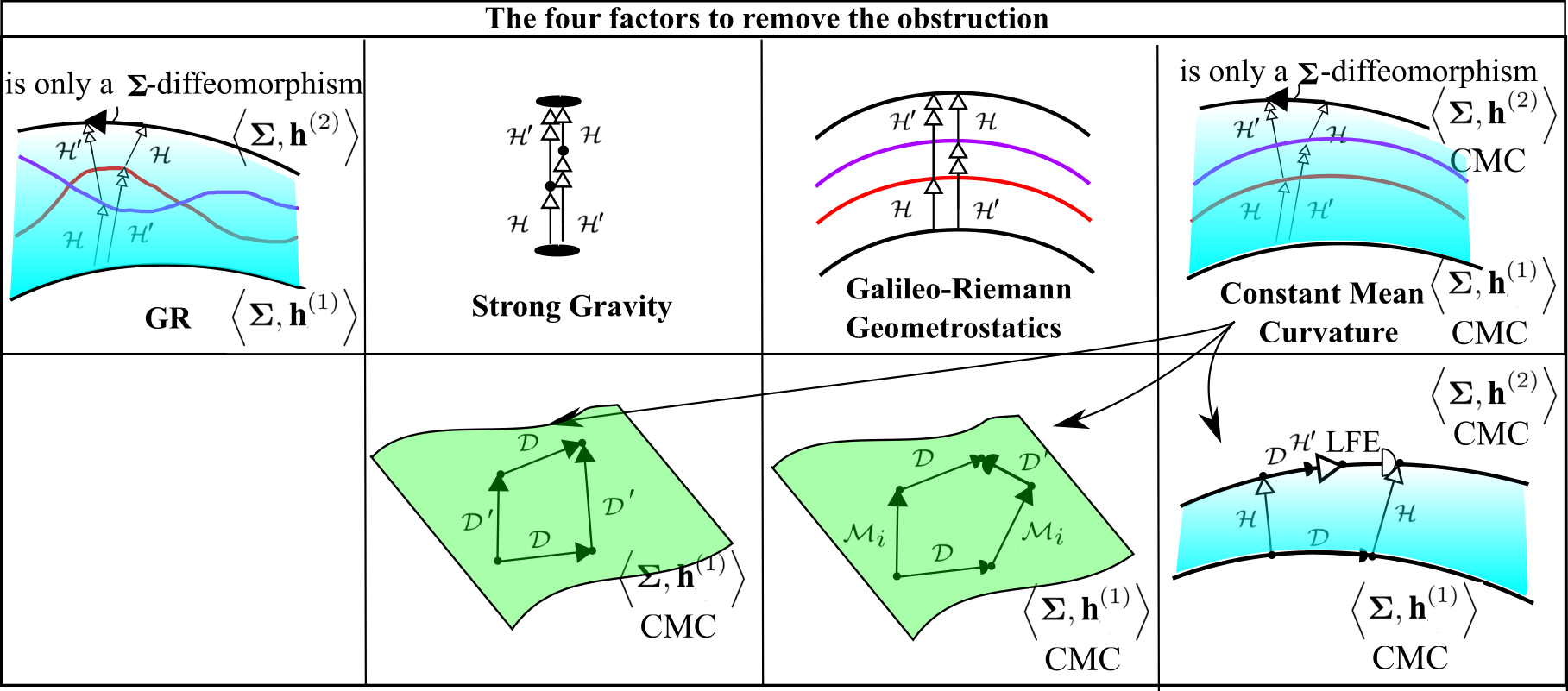}
\caption[Text der im Bilderverzeichnis auftaucht]{\footnotesize{Pictorial interpretation 
of the various brackets resulting from the current Section's geometrodynamical ansatz.
The black and white semicircle arrows indicate action of $\scD$ and a differential of the instant fixing equation respectively.   
See \cite{AForth} for the metrodynamical analogue of this Figure.}} 
\label{SCP-Figure}\end{figure}} 

\section{Strongly vanishing options: GR, Strong Gravity, Geometrostatics}\label{3-Strong}

\subsection{$\bix$ =  1 GR with embeddability into spacetime}

The third factor in (\ref{4-Factors}) strongly fixes \cite{RWR} the supermetric coefficient to $x = 1$; correspondingly, $w = 1$.  
This is indeed the DeWitt value \cite{DeWitt67} that characterizes GR [c.f.\ (II.21)]. 
In this case, GR spacetime is furthermore constructed as follows.   

\m 

\n{\bf Construction I)} The Machian version of the Thin Sandwich construct of Sec VI.7.6 applies.
One can now construct an object $\bcalC$ interpreted as an emergent object of the Machian relational form 
\beq
\frac{\d(\mbox{change})}{\d(\mbox{other change})}  \:=   \bcalC  \:=  \frac{\pa_{\sF} \bh}{\pa \mI}   \m .  
\eeq
Furthermore all the geometrical change is given the opportunity to contribute to the $\pa \mI$ that each individual change is compared to here.
So it is a STLRC entity.  

\m 

\n{\bf Construction II)} $\scH$ subsequently takes the form of the double contraction of Gauss' embedding equation that is the GR Hamiltonian constraint.  

\m 

\n Thus this matches the contraction of Codazzi's embedding equation that is the GR momentum constraint.
Consequently, a pair of embedding equations arise \cite{Tei73}.  
These constitute the 4 $0\mu$ components among the 10 components of the 4-$d$ Einstein field equations.  

\m 

\n{\bf Remark 1} The corresponding equations of motion turn out to be a linear combination of the Ricci embedding equation, 
the contracted Gauss embedding equation and the metric times further contractions. 
In this way, these equations form the TRi version of the remaining 6 Einstein field equations.

\m 

\n{\bf Remark 2} So in this approach, one {\sl recovers} equations and makes a {\sl meaningful grouping} of them. 
Contrast the decomposition of the Einstein field equations into projection equations, 
or Wheeler's suggestion of presupposing embeddability into spacetime          of \cite{Tei73, HKT} and Article XII.

\m 

\n{\bf Construction III)} One can then posit an ambient metric 4-geometry that the metric 3-geometry of space is locally embedded within.  
This could be the conventional spacetime if its signature is indefinite           alias Lorentzian: -- + + +, corresponding to $a > 0$. 

\m 

\n{\bf Remark 3} This case lies within the scope of Article XIV's Lie Mathematics applying locally.

\m 

\n{\bf Remark 4} Alternatively, it could be the counterpart whose signature is positive-definite alias Euclidean:   + + + +, corresponding to $a < 0$.
The distinction between these is not made by the Dirac Algorithm: both are consistent.  
See two Sections down for a physical dismissal of this Euclidean alternative.

\m 

\n{\bf Construction IV)} The momentum formulation is entirely unaffected by the distinction between (II.12) and (VI.72).
The ADM and relational momenta coincide in the $x = 1 = w$, $y = 0$ case for which they all exist. 
Thus in this case comparing the `ADM-momentum to $\mK_{ab}$ relation' 
and the `relational momentum to $\bcalC$ relation' permit the identification of $\bcalC$ and $2 \, \bcalK$.  
One is henceforth entitled in this $x = 1 = w$, $y = 0$ case to use the shorthand
\beq
\frac{\pa_{\sF} \bh}{2 \, \pa\mI}  \es  \bcalK                                                         \m .  
\eeq
The conventional extrinsic curvature interpretation can then be recovered by hypersurface tensor spacetime-space duality.  

\m 

\n{\bf Construction V)} At the level of the action, the relational action (VI.62) ensues from the $x = 1$ strong fixing.
This can be repackaged as, firstly, the TRi-split A action of GR, and, secondly, as the Einstein--Hilbert action of GR.
[Moreover, this end-product is a local construct in the same senses that the field equations are.]

\subsection{$\biy$ = 0 gives Geometrostatics}

\n{\bf Structure 1} In this alternative (already mentioned in Teitelboim's work \cite{Teitelboim}) the trial quadratic constraint ceases to contain a kinetic term.
This admits non-dynamical interpretation as a {\sl geometrostatics}.   

\m 

\n{\bf Remark 1} If, on the one hand, we insist that the {\sl action} must be built from first principles, however, this geometrostatics option is not possible.
Then the Relational Approach requires a geometrodynamics in which the geometry indeed undergoes nontrivial dynamics.  

\m 

\n{\bf Remark 2} If, on the other hand, we attribute primary significance to the constraint algebraic structure, 
this alternative is allowed in both such a restriction of the Relational Approach \cite{Phan} and in the Deformation Approach \cite{Teitelboim}.

\subsection{$\bia$ = 0 gives Strong Gravity} 

\n{\bf Structure 1} In this alternative, the potential ceases to contain a Ricci scalar since the cofactor of $a$ in the action is ${\cal R}$,  
For $w = 1 = x$, this amounts to recovering the Strong Gravity that corresponds to the strong-coupled limit of GR.  

\m 

\n{\bf Remark 1} Removing the above obstruction term, however, in no way requires fixing the supermetric coefficient $w$. 
Instead, a family of theories for arbitrary $w$ arises in this manner. 
This can moreover be interpreted as strong-coupled limits of Scalar--Tensor Theories that likewise apply in the vicinity of singularities in those theories.  
Clearly from each worldline only being able to communicate with itself, 
other than near singularities these geometrodynamical theories very much do not match everyday Physics.  

\m 

\n{\bf Remark 2} Henneaux's work \cite{Henneaux79} can moreover be interpreted as the hypersurface derivative  
or spatial corrected derivative maintaining \cite{FileR} 4-space to 3-space duality, with the 4-objects involved having a distinct nature from GR's.
Henneaux \cite{Henneaux79} and Teitelboim \cite{Teitelboim} followed this up by working out the Strong Gravity analogue of the geometry of hypersurfaces within spacetime.  
This turns out to have a degenerate-signature manifold (0 + + +) for its split space-time structure.   
In this way, Strong Gravity serves as an example that such duality is not exclusive to GR spacetime and its Euclidean counterpart.  

\m

\n{\bf Remark 3} It thus turns out that both Geometrostatics and Strong Gravity greatly simplify the constraint algebraic structure.   
This is because these are {\sl factors in common} with the momentum constraint arising from the Poisson bracket of two trial-Hamiltonian constraints.
In this way, $\u{\scM}$ is not an integrability of the corresponding $\scH^{\st\sr\si\sa\sll}$.  

\m 

\n{\bf Remark 4} Additionally, by strongly killing off the right hand side of the Poisson bracket of two $\scH$'s, 
the algebraic structure ceases to involve any structure functions. 
Thus it is a bona fide algebra rather than an algebroid.

\m 

\n{\bf Remark 5} All in all, (VII.49-50) and the Abelian simplification (III.56) of (VII.51) are obeyed \cite{I76, Teitelboim} in each of the last two cases. 
These correspond to entirely opposite representations of the object $\scH$: pure potential and `pure kinematical plus $\slLambda$-term' cases respectively.

\subsection{Avoiding GR's integrability}

\n{\bf Geometrodynamical Integrability Theorem}. 
If the geometrodynamical ansatz (\ref{trial}) is assumed, then the following three outcomes are consistent.  

\m 

\n i)  Recovery of GR-as-Geometrodynamics' integrability condition \cite{MT72}.   

\m 

\n ii) Geometrostatics free from any integrability condition.

\m 

\n iii) A 1-parameter family of strong gravity theories free from any integrability condition.

\m 

\n{\u{Proof}} (\ref{Int}) has 3 factors 
\be 
y \times a \times \u{\scM}
\ee 
corresponding to these three alternatives. $\Box$ 

\m 

\n{\bf Corollary} In Strong Gravity \cite{San} and Geometrostatics, Temporal and Configurational Relationalism can be independently implemented. 

\m 

\n{\bf Remark 1} This is the same situation as in RPM, but not in GR, where Temporal Relationalism enforces Configurational Relationalism as an integrability.

\section{Family ansatz for addition of minimally-coupled matter}\label{Matter}

This extension is required for the next Section's consideration of the local Relativities corresponding to each option.
This is a requirement from the following perspective.  
These local Relativities are not a property of some container spacetime but rather of all the physical laws bar Gravitation 
(which is less local as per Chapter 6 of \cite{ABook}). 
Thereby, the framework requires extension to include at least two matter field laws.

\m 

\n{\bf Structure 1} To this end, we introduce fundamental bosonic matter fields $\uppsi^{\sfA}$ of unspecified tensorial rank; 
it turns out that a sufficient set of these can be treated all at once.
These are as per Section VI.9 but with split-off species-wise coefficients $y_{\uppsi}$ and $a_{\uppsi}$, 
We use 
$$
\pa \ms^{\sg\sr\sa\sv-\uppsi} = \sqrt{\pa \ms^{\sg\sr\sa\sv \, 2}_{y, w} + \pa \ms^2_{\uppsi}} \mbox{ with } \m  
\pa \ms_{\uppsi}^2 := \sum_{\sfz \in \sfZ}y^{-1}_{\uppsi}\mM_{\sfz\sfz^{\prime}}\pa \uppsi^{\sfz}\pa \uppsi^{\sfz^{\prime}}
$$ 
for configuration space metric $\mM_{\sfz\sfz^{\prime}}$ blockwise corresponding each species $\uppsi^{\sfz}$. 
This is taken to be ultralocal in the spatial metric $\bh$ and with no dependence on the matter fields themselves.

\m 

\n{\bf Structure 2} We also adopt the potential factor 
$$
{\cal W}^{\sg\sr\sa\sv-\uppsi} := a \, {\cal R} + b + \sum_{\uppsi}a_{\uppsi} \mU_{\uppsi} \m .
$$  
This can only depend on the spatial derivatives of the spatial metric through the spatial Christoffel symbols.  

\m 

\n{\bf Structure 3} For many purposes an equivalent starting point is 
$$
\scH_{x, y, y_{\uppsi}, a, a_{\uppsi}, b}  \:=  {||\bp||_{\sbN^{x, y}}}^2  + {||\Pi||_{\sbN_{\uppsi}}}^2 - \overline{a\,{\cal R} + b + \sumpsi a_{\sfz} \mU_{\sfz}} = 0  \m .
$$
\be 
\:=  \mN_{abcd}^{x,y}\mp^{ab}\mp^{cd} + \sumpsi y_{\uppsi} \mN_{\uppsi}^{\sfz\sfz^{\prime}}\Pi_{\sfz}\Pi_{\sfz^{\prime}}/\sqrt{\mh} 
                                                                                                        - \overline{a\,{\cal R} + b + \sumpsi a_{\sfz} \mU_{\sfz}} = 0  \m .						 
\eeq 
\n{\bf Remark 1} For these models, changes in all the matter degrees of freedom do have the opportunity to contribute to  
\beq
\lt^{\se\sm}_{\nFrg\mbox{\scriptsize -free}}  \es  \int \frac{  \pa \ms^{\sg\sr\sa\sv-\uppsi}  }{  \sqrt{  2 \overline{W}^{\sg\sr\sa\sv-\uppsi}  }  } \m .  
\eeq
\n{\bf Proposition 1} The new Poisson bracket of $\scH_{x, y, y_{\uppsi}, a, a_{\uppsi}, b}$  with $\scM_i^{\sg\sr\sa\sv-\uppsi}$ 
is the obvious result of a 3-diffeomorphism Lie dragging.

\m 

\n{\bf Proposition 2}
$$
\mbox{\bf \{} (    \scH_{x, y, y_{\uppsi}, a, a_{\uppsi}, b}    \, | \,    \pa \mJ    ) \mbox{\bf ,} \, (    \scH_{x, y, y_{\uppsi}, a, a_{\uppsi}, b}    \, | \,    \pa \mK    ) \mbox{\bf \}} =  
\left(
ay \{  \scM_i^{\sg\sr\sa\sv-\uppsi} + 2 \{1 - x\} {\cal D}_i \mp \} 
\right.
$$
$$
\left.
+ \sumpsi
\left\{ 
{ a y
\left\lceil
\Pi^{\sfZ}\frac{\updelta\pounds_{\pa{\underline{\sL}}} \uppsi_{\sfZ}}{\updelta\pa{\mL}^i}
\right\rceil } 
       \left.
- 2 \, a_{\uppsi}y_{\uppsi} {        M^{\sfZ\sfZ^{\prime}} \Pi_{\sfZ}  \frac{    \pa {\cal U}_{\uppsi}    }{   \pa \, \pa_i\uppsi^{\sfZ^{\prime}})   }        }    
      \right\}
	  \right.
	  \right.
$$	   

\n \beq   
\left.
\left.
	     - 2 { y        
			\left\{
 \mp_{jk} - \frac{x}{2}\mp \, \mh_{jk}
            \right\}
\mh_{il}
\sumpsi   a_{\uppsi}            
			\left\{
            \frac{    \pa \mU_{\uppsi}    }{    \pa {\Gamma^c}\mbox{}_{jl}    }\mh^{ck} - \frac{1}{2}\frac{    \pa {\cal U}_{\uppsi}    }{    \pa{\Gamma^c}\mbox{}_{jk}    }\mh^{lc}
            \right\}    }
\right| 
\pa \mJ \, \overleftrightarrow{\pa}^i  \pa \mK 
\right)
\m .
\label{4-Factors-Matter}
\eeq
\n{\bf Remark 3} The third term's `ceiling parenthesis' denotes the extent to which the variational derivative inside acts. 

\m 

\n{\bf Remark 4} The above-listed matter fields all have no Christoffel symbol terms in their potentials, so the last underlined grouping drops out.

\section{The 3 strong obstruction factors as Relativities}

\subsection{The GR case of Geometrodynamics possesses locally Lorentzian Relativity}

\n{\bf Remark 1} Here 
\be
a y = a_{\sfz} y_{\sfz}
\ee 
arises, by which matter wave equations are formed between the first and second underlined terms. 
In this way, 
\be 
c_{\sfz} = c_{\sg\sr\sa\sv}
\ee 
is enforced: each of these matter fields $\uppsi^{\sfz}$ is forced to have the same maximum propagation speed $c_{\sm\sa\sx}$   
-- and consequently null cone -- as Gravitation.
Thereby, any pair of these matter fields $\uppsi^{\sfz}$, $\uppsi^{\sfz^{\prime}}$ are forced to share these entities with each other: 
\be 
c_{\sfz} = c_{{\sfz^{\prime}}}
\ee 
and a common null cone for $\fz$ and $\fz^{\prime}$.
In this way, the Relational Approach {\sl derives} rather than assumes the Lorentzian Relativity Principle, as a consistency condition \cite{RWR, AM13} 
corresponding to the universal local cone in Fig \ref{Gal-Lor-Ca-Causality}.b).

\m 

\n{\bf Remark 2} The Euclidean-signature case which also arises in this manner does not occur physically 
as is clear from the observed existence of finite propagation speeds.

\subsection{Geometrostatics possesses locally Galilean Relativity}

\n In this case, the shared maximum propagation speed
\be 
c_{\sm\sa\sx} = \infty   \m .  
\ee 
This amounts to the local SR null cones have been squashed into planes, which is the Galilean limit of causal structure: Fig \ref{Gal-Lor-Ca-Causality}.a).  
In the flat-space case, this amounts to a derivation of Galilean Relativity, 
in fact of an in-general curved-space geometrostatics which is a `Galileo--Riemann' generalization \cite{AM13}.

\subsection{Strong Gravity geometrodynamics possesses locally Carrollian Relativity}.   

\n In this case, the shared maximum propagaion speed
\be 
c_{\sm\sa\sx} = 0         \m . 
\ee
Thus the null cones become squeezed into lines, so that each point can only communicate with its own worldline.
This consequently possesses Carrollian Relativity: Fig \ref{Gal-Lor-Ca-Causality}.c). 
Henneaux \cite{Henneaux79} pointed out that Strong Gravity exhibits this limit of null cones.
Klauder \cite{Klauder70} additionally studied such a zero propagation speed matter sector: `ultralocal matter fields'.  

\m

\n{\bf Remark 1} In these last two cases (some of) the matter can be the opposite limiting case to the gravitational sector.  
Of course, none of the options in this paragraph are physically realistic.   
%
{\begin{figure}[!ht]
\centering
\includegraphics[width=1.0\textwidth]{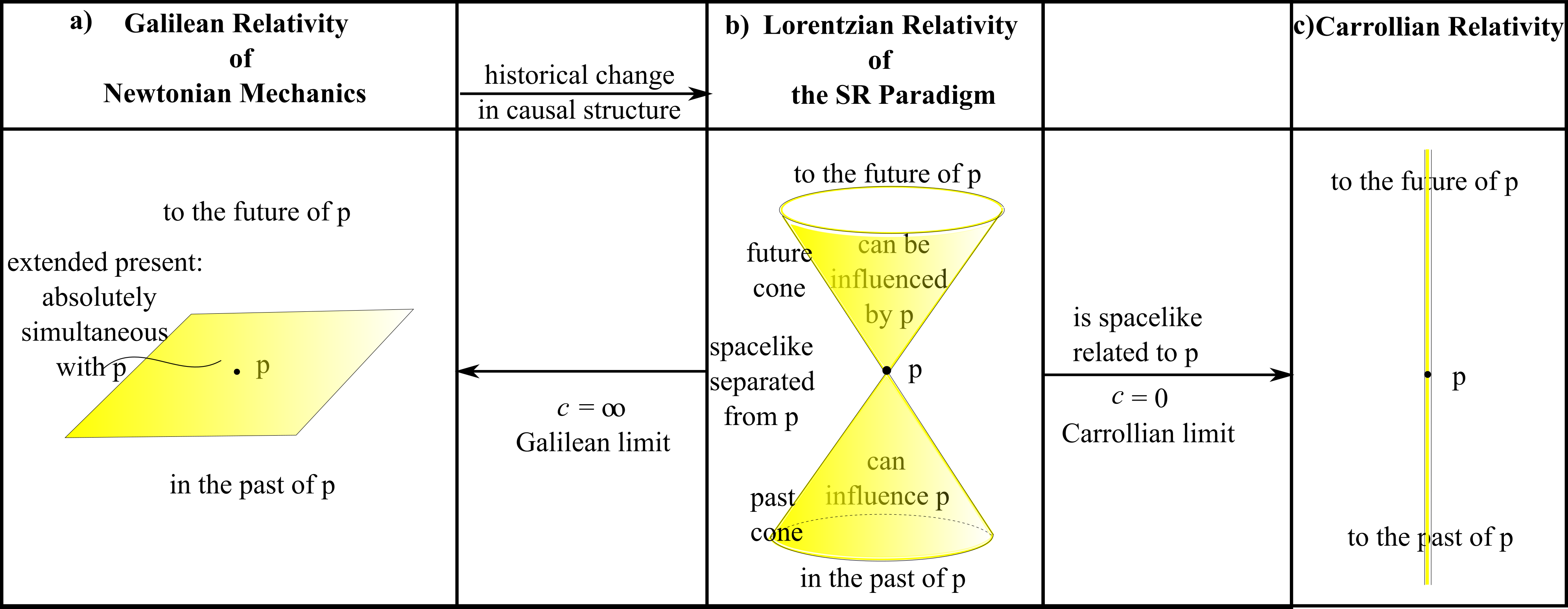}
\caption[Text der im Bilderverzeichnis auftaucht]{\footnotesize{
a) \index{past}Past, present\index{present} and future\index{future} of an event $p$ in Newtonian Mechanics.
b) Past and future null cones\index{null cone} of an event $p$ in Minkowski spacetime $\mathbb{M}^4$.\index{spacetime!flat -}  
a) is the      Galilean       limit of b) in which the null cone is squashed into a plane.
c) is the opposite Carrollian limit of b) in which the null cone is squeezed into a line.}} 
\label{Gal-Lor-Ca-Causality}\end{figure}} 

\subsection{Discussion}

\n{\bf Remark 1} As a package, the three possible strong ways of evading the obstruction term are the 
\be
\mbox{maximum propagation speed } \m  c_{\sm\sa\sx} = 0 \mma \mbox{finite and } \m \infty \m \mbox{ trichotomy} \m .  
\ee  
Local Relativity now follows from closure of the constraint algebroid rather than being postulated a priori as it was historically by Einstein. 
Physical observation of finite maximum propagation speed serves to select GR alone out of the above set of theories.  

\m 

\n{\bf Remark 2} Suppose moreover that one adopts the physical choice of locally Lorentzian Relativity. 
This comes hand in hand with {\sl deducing} embeddability into a GR spacetime, a formalism and worldview which has long been known to 
be widely insightful \cite{HE, Wald}. 

\m

\n{\bf Remark 3} This Section can furthermore be interpreted as an answer to Wheeler's question (II.25).  
So the GR form of $\scH$ arises as one of but a few consistent options upon assuming just the structure of space.
This answer ascribes primality to space rather than to spacetime and yet leads to spacetime emerging.  
Thus it provides a resolution of the classical Spacetime Construction Problem as well, in the sense of construction from space.
This result can be considered to arise from local SR, GR and its spacetime structure being rigid, rather than mutable as functioning mathematical structures.

\subsection{Comparison with Einstein's historical route to GR}

{\bf Remark 1} On the one hand, Einstein chose to consider spacetime primality instead of spatial primality. 
In this setting, he changed the status of frames from SR's Lorentzian inertial frames to local inertial frames that are freely falling frames.  
This made direct use of the spacetime connection in passing locally to freely falling frames.    

\m 

\n{\bf Remark 2} Considering the corresponding curvature tensors is now natural, 
and leads to a law relating the Einstein curvature tensor to the energy--momentum--stress tensor of the matter content.  
This accounts for the local inertial frames on physical grounds, and the spacetime geometry is to be solved for rather than assumed. 
This approach does not however directly address Machian criteria for time and space.  

\m 

\n{\bf Remark 3} On the other hand, in the Relational Approach, the horn in which space is primary is chosen.
Time and space are conceived of separately, and Leibniz--Mach relational criteria (as per Articles I and II) are directly applied to each. 
The notion of space is broadened from that of the traditional absolute versus relational debate arena so as to include geometrodynamical theories. 
This approach's equations pick out a particular subcase for which an ambient spacetime manifold is implied; 
this is a very fruitful perspective as per Chapter 6 of \cite{ABook}.  

\m 

\n{\bf Remark 4} So in the Relational Approach, SR arises as an idealization that holds well locally in many parts of the Universe.  
SR's assumed universal symmetry group and shared null cone is explained in the Relational Approach as emergent phenomena.
%

\m 

\n{\bf Remark 5} This formulation having constructed spacetime curvature from its split space--time form, 
it is conversely now natural to ask whether spacetime connections also play a role in the theory: a `Discover Curvature and then Connections' approach.  

\m 

\n{\bf Remark 6} On the one hand, one can build the latter out of elements natural to the spatially primary perspective as per Article XII. 
%
%
Thus in the Relational Approach one finally arrives indirectly at the identification of local inertial frames with freely falling frames. 
%

\m 

\n{\bf Remark 7} On the other hand, in the spacetime formulation of GR, one has the Equivalence Principle modelled by connections prior to bringing in curvature: 
a `Discover Connections and then Curvature' approach.  

\m 

\n{\bf Remark 8} As regards a few alternative theories of gravity, 
the final version of \cite{RWR} showed that Brans--Dicke Theory can also be cast in relational form. 
This extension contains metric--matter cross-terms.
%
%
Different values of the Brans--Dicke parameter furthermore become consistent via involvement of metric--matter cross-terms. 
In this way, Brans--Dicke theory and other more complicated Scalar--Tensor Theories are available not only as a strong gravity 
limit resolutions of Spacetime Construction but also as finite propagation speed alternatives to the GR outcome \cite{San}. 

\m 

\n{\bf Remark 9} Inclusion of fermions \cite{Van, FileR} requires a linear kinetic term ${\cal T}_{\sll\si\sn}$ 
being additively appended to the product of square roots as per Sec VI.9.

\section{The fourth weakly-vanishing factor}\label{1-weak}

(\ref{4-Factors})'s fourth factor contains a $\bcalD \, \mp$ core, the vanishing of which can be written as 
\beq 
\bcalD \left\{   \frac{\mp}{\sqrt{\mh}}  \right\} = 0
\label{Dp}
\eeq 
without loss of generality, since $\sqrt{\mh}$ is covariantly constant.
(\ref{Dp}) can be formulated as
\beq
\scD := \mp - \sqrt{\mh} \, c = 0
\label{CMC2}
\eeq
for $c$ spatially constant, by performing one integration. 
This presents a weakly-vanishing option, covering maximal slices 
\be 
\mbox{\it maximal slices } \m \mp = 0
\ee 
and, more generally, 
\be 
\mbox{\it constant mean curvature slices } \m \frac{\mp}{\sqrt{\mh}} = const   \m .  
\ee 
These considerations eventually motivate more general realizations of solutions than the above; see e.g.\ \cite{MBook}.

\section{Discover-and-encode approach to Physics}\label{DAE}

At the classical level, this amounts to trying out a $\lFrg$, 
finding it gives further integrability conditions that enlarge $\lFrg$ and then deciding to start afresh with this enlarged $\lFrg$.

\subsection{Metrodynamics assumed ($\lFrg$ = id)}  

\n This is a more minimalist assumption \cite{OM02, San} than assuming a geometrodynamics.  
It is a simple Comparative Background Independence result, taking the following form. 

\m 
 
\n{\bf Consistent Metrodynamics Theorem}.  Consider ansatz (\ref{trial}) but with $\lFrg$ = id, so the $\pa \underline{\mF}$ are removed.  
Then the following five outcomes are consistent.  

\m

\n i) Recovery of GR, through $\scM_i$ appearing as an integrability condition thus forcing $\lFrg$ = id's enlargement to $Diff(\bigupsigma)$.

\m 

\n ii)  A 1-parameter family of metrostatics.

\m 

\n iii) A 1-parameter family of Strong Gravity metrodynamics theories.

\m 

\n iv)  A group of formulations and theories based upon 
\be 
\bcalD \left\{\frac{\mp}{\sqrt{\mh}}\right\} = 0   \m .
\ee 
\n v)   A `unit-determinant geometrodynamics',  
corresponding to \cite{AM13} $\lFrg$ = id's enlargement to the group of unit-determinant diffeomorphisms, 
\be 
UDiff(\bigupsigma)
\ee 
\underline{Derivation} \cite{OM02, San, AM13} In this case, to start off with there is just a primary constraint $\scH_{x,y,a,b}$.  
In considering the Poisson brackets this forms with itself, one no longer has an initial right to a priori `parcel out' an $\scM_i$. 
One is instead to use the first form of the right-hand side of (\ref{Htrial-Htrial}) and define
\beq
\u{\scS}^x :=  2 \{ -\u{\bcalD} \, \u{\u{\bp}}  + \{1 - x\} \u{\bcalD} \,  \mp \}
\label{Daenerys} 
\eeq 
as the preliminary secondary constraint entity arising from this Poisson bracket.  
This is now smeared with some differential vector $\pa \bupiota$.  

\m 

\n (\ref{Daenerys}) features unless one of $a = 0$ or $y = 0$ holds, in which case the above right hand side strongly vanishes. 
The Abelian constraint algebra of Sec VII.5.2 applies in both of these cases. 
Moreover, each case involves a diametrically opposite representation of the $\scH$ object itself. 
I.e.\ the mostly kinetic $\scH_{0,b,x,y}$ of Galileo--Riemann metrostatics: case ii) versus the zero-kinetic term $\scH_{a,b,x,0}$ of strong metrodynamics: case iii).   

\m 

\n[These theories are not the same as the previous Section's, 
since now they have no diffeomorphism-related constraints and thus remain metrodynamical rather than geometrodynamical theories.]

\m 

\n If $\bscS$ is present, 
$$
\mbox{\bf \{} (    \sbcS    \, | \,    \pa \bupiota    ) \mbox{\bf ,} \, (    \sbcS  \,  | \,  \pa \bupchi^j   ) \mbox{\bf \} }                               =  
                       \big(    - 2 \, \u{\bcalD} \u{\u{\bp}} + 2 \{1 - x\} \{3x -  2 \} \u{\bcalD} \mp   \,  \big| \, \u{[\pa \upiota,\pa \upchi]} \big) \m  =
$$
\be
	            		     \big(    - 2 \, \u{\bcalD} \u{\u{\bp}} \big| \, \u{[\pa \upiota,\pa \upchi]}    \big)  
+ 2   \{1 - x\} \{3x -  2 \} \big(           \u{\bcalD} \mp 	    \big| \, \u{[\pa \upiota,\pa \upchi]}    \big) \,  
\label{lf}
\ee
ensues. 
Comparing (\ref{Htrial-Htrial}) and (\ref{lf}) gives that the constraint algebroid closes only if $x = 1$, $x = \frac{2}{3}$ or ${\bcalD} \mp = 0$.  
The last of these gives case iv) as usual.

\m 

\n If $x = 1$, then $\bscS^x$ collapses to $\u{\scM}$ the generator of diffeomorphisms and therefore the main GR case i) of the working is recovered.

\m 

\n If $x = \frac{2}{3}$ instead,  a distinct clear geometric meaning arises as follows.  
The corresponding 
\beq
\u{\bscU} \:=   \u{\bscS}^{2/3}  
          \es  - 2 \left\{ \u{\bcalD} \, \u{\u{\bp}}  - \mbox{$\frac{1}{3}$} \, \u{\bcalD} \, \mp   \right\}  \m 
\eeq
is the generator of {\it unit-determinant diffeomorphisms}: diffeomorphisms that preserve the local volume element $\sqrt{\mh}$: case v). $\Box$  

\m 

\n{\bf Remark 1} At the level of $\Riem(\bigupsigma)$, this corresponds to picking the degenerate (null signature) supermetric.  
This degeneracy means that case v) has no underlying relational action.  
This theory's exact meaning remains unknown; 
it is an example of a theory lying somewhere between a metrodynamics and a geometrodynamics. 
We introduce the names $U$-diffeomorphism, $U$-geometry and $U$-superspace for this theory's counterparts of the geometrodynamical entities.  
$U$-diffeomorphisms use up 2 degrees of freedom per space point, leaving $U$-geometry with 4. 

\m 

\n{\bf Remark 2} In this Series, we pass on considering conformal options in any detail (see e.g.\ \cite{ABFO, ABFKO, MBook}).

\section{Brackets Consistency from Polynomial ans\"{a}tze in Flat Geometry}

We here posit polynomial ans\"{a}tze, 
and rely on just Lie brackets consistency to extract geometrical automorphism groups that are suitable for supporting theories of Geometry.

\subsection{Preliminary group notation}

\n A somewhat more extensive set of groups is required for the current section.  
In dimension $d$, $Tr(d)$ are translations and $Rot(d)$ are rotations. Dilations $Dil$ are dimension-independent. 

\m 

\n $Dilatat(d)$ are dilatations: translations and dilations. 

\m 

\n $Eucl(d)$ are Euclidean transformations: translations and rotations. 

\m 

\n $Sim(d)$ are similarities: translations, rotations and dilations.  

\m 

\n $Aff(d)$ are affine transformations: similarities alongside shears and Procrustes stretches; 
this can also be viewed as general-linear transformations alongside translations.

\m 

\n $Proj(d)$ are projective transformations: affine transformations alongside special-projective transformations. 

\m

\n $Conf(d)$ are conformal transformations: similarities alongside special-conformal transformations.

\m

\n The prefixes $P\mbox{-}Iso\mbox{-}$ and $C\mbox{-}Iso\mbox{-}$ denote the result of replacing translations by 
special-projective and special-conformal transformations respectively.

\subsection{1-$d$ case}

Let us first carry this out in this simpler case.  

\m 

\n{\bf Preliminary Lemma} i) Assuming $\d_x$ returns $Tr(1) = Eucl(1)$. 

\m 

\n ii) $\langle x \, \d_x \rangle$ returns $Dil(1)$.  

\m

\n iii) $\langle \d_x, \, x \, \d_x \rangle$ returns $Dilatat(1) = Sim(1)$. 

\m

\n iv) $x^2 \d_x$ returns $P\mbox{-}Iso\mbox{-}Tr(1)$.   

\m

\n v) $\langle x \, \d_x, \, x^2 \d_x\rangle$ returns $P\mbox{-}Iso\mbox{-}Dilatat(1)$.   

\m

\n vi) $\langle \d_x, \, x^2 \d_x \rangle$ fails to close alone. 
\n vii) $\langle \d_x , \, x \, \d_x, \, x^2 \d_x \rangle $ gives $Proj(1)$. 

\m

\n viii) Trying to extend using any cubic or higher terms 
\be 
x^{n} \d_x \mma n \geq 3 
\ee 
produces a cascade.  

\m

\n{\bf Remark 1} Guggenheimer pre-empted results vii) and viii) \cite{G63}.  
He also gave ii) and iv)'s generators. 
We go further by considering the two of these together. 
By associating $P\mbox{-}Iso$ gometries in their own right to iv) and v). 
And by reinforcing the cascade with a counting argument precluding $N$-point Shape Theory 
(as well as $N$-point invariants: a notion that Guggenheimer -- and \'{E}lie Cartan \cite{Cartan55} -- already possessed).  

\m 

\n{\bf Remark 2} I since found that vii) and viii) can furthermore be readily inferred from Chapters 3 and 4 of Lie's 1880 treatise \cite{Lie80}.  

\m 

\n{\bf Remark 3} vi) reflects the 1-way mutual integrability
\be 
(P, Q) \, \, \Thomas \, \, D                           \m .  
\label{PQ->D}
\ee

\subsection{Higher-$d$ case}

Let us now go further than Guggenheimer by considering higher-dimensional ans\"{a}tze \cite{A-Brackets}.
The general (bosonic vectorial) quadratic generator in $\geq 2$-d is the following 2-parameter family ansatz
\be 
\u{Q}^{\st\sr\si\sa\sll}_{\mu, \nu}  \:=  \mu ||x||^2 \u{\pa}  +  \nu \, \u{x} (\u{x} \cdot \u{\nabla} )                \m .  
\ee 
This follows from considering the general fourth-order isotropic tensor contracted into a symmetric object $x^Ax^B$.  

\m 

\n{\bf Theorem} For $d \geq 2$, $\u{Q}^{\st\sr\si\sa\sll}_{\mu, \nu}$ self-closes only if either 
\be 
\mu = 0 
\label{P}
\ee 
or 
\be 
\nu = - 2 \, \mu                                                                                                             \m . 
\label{C}
\ee 

\n {\bf Remark 1} The first factor vanishing is a recovery of the special-projective generator 
\be 
\u{Q} =  x_A (\u{x} \cdot \u{\nabla} )                     \m , 
\ee 
whereas the second is a recovery of the special-conformal generator            
\be 
\u{K} =  ||x||^2\pa_A - 2 \, x_A (\u{x} \cdot \u{\nabla})  \m . 
\ee 
In this manner, the conformal versus projective alternative for flat-space top geometry is recovered.  			   

\m 

\n{\u{Proof}} See \cite{A-Brackets}. The key line is
\be 
\mbox{\bf [} Q^{\st\sr\si\sa\sll}_{\mu, \nu \, A} \mbox{\bf ,} \, Q^{\st\sr\si\sa\sll}_{\mu, \nu \, B} \mbox{\bf ]}  \es  2 \, \mu \, (2 \mu + \nu) ||x||^2 x_{[A}\pa_{B]}   \m . 
\ee
\n{\bf Corollary 1} Each of $\u{Q}$ and $\u{K}$ is self-consistent, returning $P\mbox{-}Iso\mbox{-}Tr(d)$ 
                                                                          and $C\mbox{-}Iso\mbox{-}Tr(d)$ respectively.

\m 

\n The following supporting Lemma enables a number of further Corollaries. 

\m 

\n{\bf Supporting Lemma} i) The general zeroth order generator $\u{P} := \u{\pa}$ is self-consistent according to (\ref{P-P}), returning $Tr(d)$.  

\m 

\n ii) The general scalar and 2-tensor linear ans\"{a}tze      
\be 
D := \u{x} \cdot \u{\pa} \m \mbox{ and }
\ee  
\be 
G_{AB} := x_A \pa_B
\ee 
are each self-consistent according to 
\be 	 
\mbox{\bf [}  D   \mbox{\bf ,} \, D  \mbox{\bf ]}                    =  0                                                                             \m , 
\label{D-D}
\ee 
and 
\be 	 
\mbox{\bf [}  {G^A}_B \mbox{\bf ,} \, {G^C}_D  \mbox{\bf ]}  =  2 \, {\delta^{[C}}_B {G^{A]}}_D                                                       \m ,  
\label{G-G}
\ee 
returning $Dil(d)$  and the general-linear group $GL(d, \mathbb{R})$ respectively \cite{A-Brackets}.  

\m 

\n In fact, for $d \geq 2$, ansatz $D$ is redundant since this the trace part of $G_{AB}$: 
\be 
{G^A}_A = \u{x} \cdot \u{\pa} = D   \m ,  
\ee 
whereas $d = 1$ exhibits the coincidence 
\be 
G  \es  x \, \d_x 
   \es      D                       \m .  
\label{G-Def-1-d}
\ee
\n iii) These zeroth- and first-order ans\"{a}tze $\u{P}$ and $\u{\u{G}}$ are additionally mutually consistent as per 
\be 
\mbox{\bf [}  P_A \mbox{\bf ,} \, P_{B}  \mbox{\bf ]}                =  0                                                                             \m , 
\label{P-P}
\ee 
and (\ref{G-G}), returning the affine algebra $Aff(d)$ \cite{A-Brackets}.  

\m 

\n The preceding coincidence and 

\be 
\mbox{1-$d$'s lack of room for any antisymmetry } \m \Omega = 0 
\label{No-Anti}
\ee 
mean that $Aff(1)$ reduces to $Sim(1)$ and further to $Dilatat(1)$.

\m

\n{\bf Corollary 2} The $\u{Q}$ emerging from the Theorem's first strongly vanishing root 
is mutually compatible with both the zeroth- and first-order ans\"{a}tze $P$ and $G_{AB}$, 
by which $Proj(d) = PGL(d + 1, \mathbb{R})$'s algebra emerges; this is the $(d + 1)$-dimensional real projective linear group.

\m 

\n $\u{Q}$ can be combined with just $\u{\u{G}}$ with components 
\be 
G_{AB} := x_A \pa_B  \m , 
\ee                        
giving $P\mbox{-}Iso\mbox{-}Aff(d)$, with just its trace part 
\be 
D = \u{x} \cdot \u{\nabla}  \m ,
\ee                   
returning $P\mbox{-}Iso\mbox{-}Dilatat(d)$, 
with just its antisymmetric part $\u{\u{\Omega}}$, producing $P\mbox{-}Iso\mbox{-}Eucl(d)$, 
or with both of these,                              yielding $P\mbox{-}Iso\mbox{-}Sim(d)$.    
It can also be combined with just $\u{\u{G}}$'s tracefree part $\u{\u{\Sigma}}$, with components 
\be 
\sigma_{AB}  \:= 2  \, x_{(A}\pa_{B)}  \m - \m  \frac{1}{d} \, \delta_{AB} D                                                                            \m ,  
\label{S-Def}
\ee 
\n In 1-$d$,  by (\ref{G-Def-1-d}, \ref{No-Anti}), the first, second and fourth of these conflate to $P\mbox{-}Iso\mbox{-}Dilatat(1)$, 
                                                              whereas the third and fifth conflate to $P\mbox{-}Iso\mbox{-}Tr(1)$.  

\m 

\n $\u{Q}$ cannot however be combined with the tracefree symmetric part $\u{\u{\sigma}}$ in the absense of the antisymmetric part $\u{\u{\Omega}}$, 
by the 1-way self-integrability 
\be 
\u{\u{\sigma}} \, \, \Thomas \, \,  \u{\u{\Omega}}                                                                                                                \m .
\label{S->O}
\ee  
\n Finally, $\u{Q}$ cannot be combined with the zeroth-order ansatz $\u{P}$ in the absense of the linear ansatz $\u{\u{G}}$ or of any irreducible part thereof, 
by the 1-way mutual integrability 
\be 
(\u{P}, \u{Q}) \, \, \Thomas \, \, \u{\u{G}}   \m . 
\label{PQ->G}
\ee 
\n{\bf Corollary 3} The $\u{K}$ emerging from the previous Theorem's second strongly vanishing root is mutually compatible with the zeroth-order ansatz 
                        $\u{P}$ alongside the first-order ansatz's trace and antisymmetric parts $D$ and $\u{\u{\Omega}}$. 
In this way, the relation 
\be 
Conf(d)  \es  SO(d + 1, 1) \m : \m \mbox{ the $d + 1$ dimensional Lorentz group}  
\label{Conf(d)}
\ee 
emerges for $d \geq 3$, and the $SO(3, 1)$ subalgebra of the infinite $Conf(2)$ algebra emerges for $d = 2$.  

\m 

\n $\u{K}$ cannot however be combined with the symmetric part of the linear ansatz by the cascade-sourcing obstruction 
\be
\mbox{\bf [} K_A \mbox{\bf ,} \, \sigma_{BC} \mbox{\bf ]}   \es     2 \, \delta_{A(B} K_{C)}   \m + \m \frac{1}{d} \, \delta_{BC} K_A  \m + \m 
                                                                    8 \, \delta_{A(B} Q_{C)}   \m - \m \frac{1}{d} \, \delta_{BC} Q_A  \m - \m  4\, R_{A(BC)}  \m ,   
\label{K-S-Obs}
\ee
\n $\u{K}$ can be combined with just $D$,    giving $C\mbox{-}Iso\mbox{-}Dilatat(d)$, 
              with just $\u{\u{\Omega}}$, returning $C\mbox{-}Iso\mbox{-}Eucl(d)$, 
                   or with both of these,  yielding $C\mbox{-}Iso\mbox{-}Sim(d)$.    

\m 

\n{\bf Remark 2} Via some parts of the Supporting Lemma and the first part of each of Corollaries 2, and 3, 
our `Brackets Consistency' Pillar of Geometry {\sl derives} both Projective and Conformal Geometry in flat space.  
This is moreover a {\sl conceptually new type} of derivation from previous ones in the literature, 
by which a new foundational paradigm for each of Projective and Conformal Geometry in flat space.  
These moreover arise side by side as the two roots of an algebraic quadratic equation that emerges as the right-hand-side of a Lie bracket.

\m 

\n{\bf Remark 3} This method does not pick out the infinite-$d$ extension of the conformal group in 2-$d$. 

\m 

\n Our methodology just returns the {\sl finitely generated} geometrical automorphism groups. 
It is moreover logically possible for some of the infinite cascades excluded by our method 
to have significance as infinitely-generated automorphism groups, and this then so happens to be realized in the case of $Conf(d)$.

\m 

\n This case can of course be detected by the flat conformal Killing equation collapsing in 2-$d$ to the Cauchy--Riemann equations \cite{AMP}.  
The loss of this immediate deduction if one uses instead the `Lie Brackets Closure' Approach to Geometry \cite{A-Brackets}
is a first example of a price to pay for using foundations that make less structural assumptions.

\m  

\n{\bf Remark 4} Thus both `top geometries' in flat space -- conformal and projective -- arise as the 2 roots of a single algebraic equation 
for the strong vanishing of the self-bracket of the general quadratic ansatz generator.

\m 

\n{\bf Remark 5} In 1-$d$, this result fails because antisymmetric 2-forms are not supported. 
It fails moreover ab initio since there is only one rank-4 isotropic tensor in 1-$d$: the constant, 
by which our working collapses to finding the bracket of a scalar with itself, which is of course trivially zero. 
It is now clear why the theorem excludes 1-$d$, and how earlier workings in the current paper recover the outcome of the simple special case $d = 1$.   

\m 

\n{\bf Remark 6} Lie's \cite{Lie80} own systematic classification of Lie algebras only went as high as 2-$d$.  
See also \cite{Olver, Olver2} for up-to-date reviews.  

\m 

\n{\it Quid est demonstrandum} that Lie brackets rigidity, outside of Dirac's Poisson brackets rigidity, is a realized phenomenon.


\end{document}